\documentclass[journal]{vgtc}                
\ifpdf
  \pdfoutput=1\relax                   
  \usepackage{graphicx}                
  \usepackage{tikz} 
  \usepackage{subfigure}
  \DeclareGraphicsExtensions{.pdf,.png,.jpg,.jpeg} 
\else
  \usepackage{graphicx}                
  \DeclareGraphicsExtensions{.eps}     
\fi%

\graphicspath{{figures/}{pictures/}{images/}{./}} 

\usepackage{microtype}                 
\PassOptionsToPackage{warn}{textcomp}  
\usepackage{textcomp}                  
\usepackage{mathptmx}                  
\usepackage{times}                     
\usepackage{cite}                      
\usepackage{tabu}                      
\usepackage{booktabs}                  

\usepackage{amsfonts}
\usepackage{amsmath}
\usepackage{hyperref}
\usepackage{xspace}

\onlineid{0}

\vgtccategory{Research}
\vgtcpapertype{application/design study}

\newcommand{\ie}{i.e.}
\newcommand{\eg}{e.g.}
\newcommand{\biascope}{\textsc{BiaScope}\xspace}

\title{BiaScope: Visual Unfairness Diagnosis for Graph Embeddings}

\author{Agapi Rissaki\thanks{These authors contributed equally to this work.},  
Bruno Scarone\footnotemark[1], 
David Liu, 
Aditeya Pandey, 
Brennan Klein, 
Tina Eliassi-Rad, 
and Michelle A. Borkin 
}

\authorfooter{
\item Agapi Rissaki, Bruno Scarone, David Liu, Aditeya Pandey, Tina Eliassi-Rad, and Michelle A. Borkin are with the Khoury College of Computer Sciences at Northeastern University.
\item Brennan Klein and Tina Eliassi-Rad are with the Network Science Institute at Northeastern University.
 \item E-mails: \{rissaki.a, scarone.b, liu.davi, pandey.ad, b.klein, m.borkin\}@northeastern.edu and tina@eliassi.org.
}

\teaser{
  \centering
  \includegraphics[scale=0.348]{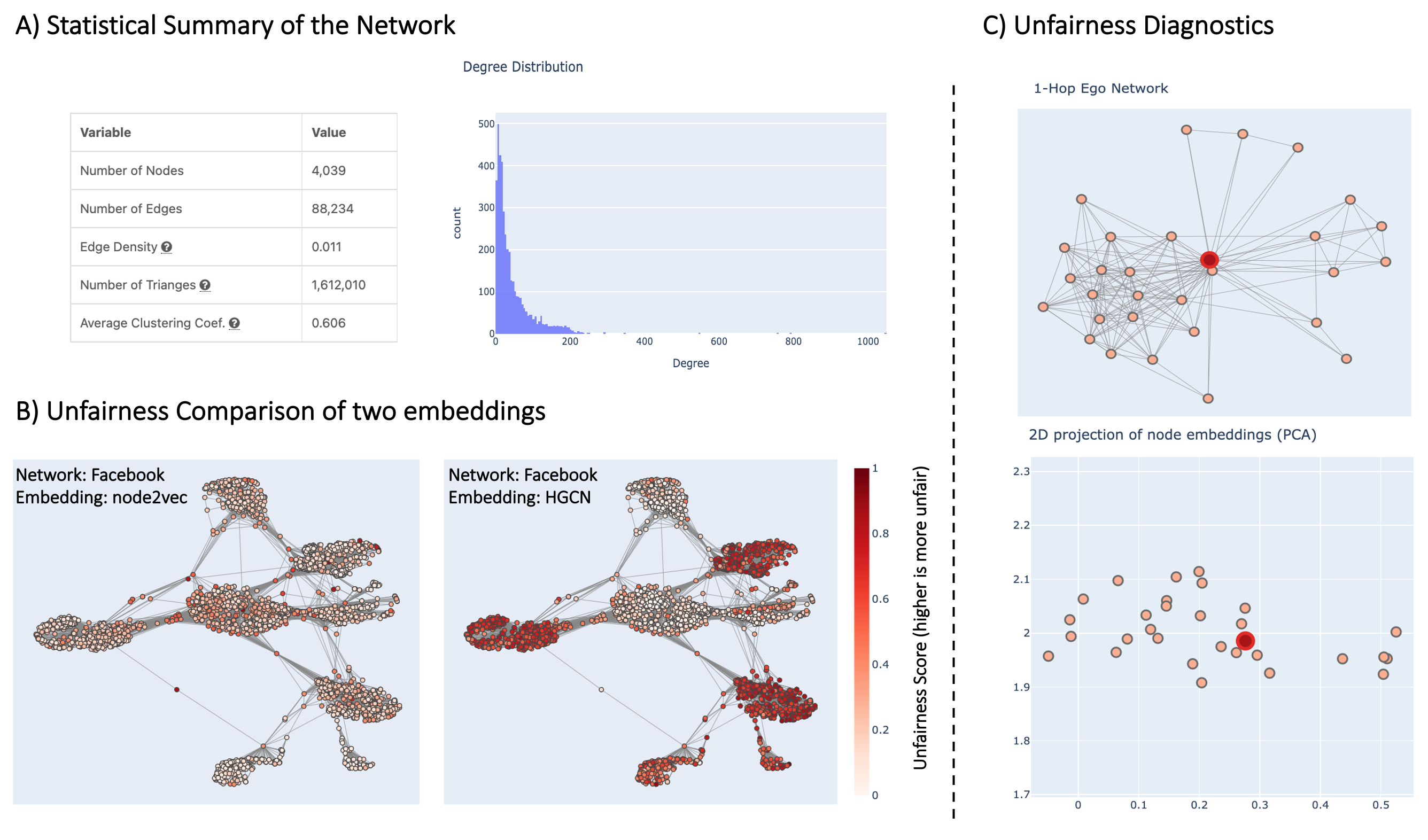} 
  \caption{\biascope is an interactive tool for \textit{visual unfairness diagnosis} for graph embeddings. The tool consists of three views: (A) a \textit{Statistical Summary View}, (B) an  \textit{Unfairness Comparison View} and (C) a  \textit{Diagnose View}. The content of the views is as follows: (A) The Statistical Summary View summarizes key properties of the selected network. (B) The Unfairness Comparison View is a side-by-side view that facilitates the fairness comparison of two embeddings of the same network. Nodes are colored based on their \textit{unfairness score}, where darker nodes are embedded more unfairly. For example, the figure shows that the observed bias can be evenly distributed throughout the network or concentrated in certain communities. (C) The Diagnose View shows the embedding subspace affecting the unfairness score of a selected node, together with its ego network. The two visualizations are interactively linked, allowing the user to investigate how bias appears in an embedding and how it relates to the relevant graph structure.}
  \label{fig:teaser}
}

\abstract{
The issue of \textit{bias} (i.e., systematic unfairness) in machine learning models has recently attracted the attention of both researchers and practitioners. For the graph mining community in particular, an important goal toward algorithmic fairness is to detect and mitigate bias incorporated into \textit{graph embeddings} since they are commonly used in human-centered applications, e.g., social-media recommendations. However, simple analytical methods for detecting bias typically involve aggregate statistics which do not reveal the sources of unfairness. Instead, visual methods can provide a holistic fairness characterization of graph embeddings and help uncover the causes of observed bias. In this work, we present \biascope, an interactive visualization tool that supports end-to-end \textit{visual unfairness diagnosis} for graph embeddings. The tool is the product of a design study in collaboration with domain experts. It allows the user to (i) visually compare two embeddings with respect to fairness, (ii) locate nodes or graph communities that are unfairly embedded, and (iii) understand the source of bias by interactively linking the relevant embedding subspace with the corresponding graph topology. Experts' feedback confirms that our tool is effective at detecting and diagnosing unfairness. Thus, we envision our tool both as a companion for researchers in designing their algorithms as well as a guide for practitioners who use off-the-shelf graph embeddings.
}


\CCScatlist{
  \CCScatTwelve{Human-centered computing}{Visualization}{Visualization design and evaluation methods}{}
  \CCScatTwelve{Theory of computation}{Design and analysis of algorithms}{Graph algorithms analysis}{}
}

\begin{document}

\firstsection{Introduction}
\maketitle





With the growing use of machine learning models to automate decisions, a rising concern for both researchers and practitioners is to assure that they are not \textit{biased} (i.e., systematically unfair) against specific population groups or individuals. More specifically, within the context of graph mining algorithms, fairness plays an important role \cite{agarwal2021towards, tsioutsiouliklis2021fairness, yao2017beyond, Kamishima12enhancementof} since these algorithms are widely used in human-centered applications with major societal impact. In graph machine learning, it is a common practice to compile network information (such as graph topology, node features) in unified node-level representations, \ie, \textit{graph embeddings}. The produced embeddings can subsequently be used off-the-shelf for a plethora of downstream tasks, \eg, node classification \cite{classification}, link prediction \cite{link}, community detection \cite{clustering}, ranking \cite{recommend}, among others. For example, graph embeddings of social networks can serve as representations of users for high-impact real-world tasks, \eg, user classification for targeted advertising, social group detection, friendship recommendation, advanced user search or to better understand a rumor spreading process \cite{hamilton2017representation, cai2018comprehensive, zhang2018network}. However, graph embeddings can carry bias against particular groups or individuals, which is either present in the data or introduced by the algorithm itself \cite{baeza2018bias}. As a result, it is crucial to identify, diagnose and mitigate bias before it gets propagated to downstream tasks and real-world machine learning systems. 

Thoroughly examining a graph embedding algorithm with respect to unfairness is hard to accomplish using solely statistical methods. While it is possible to quantify bias either at a global level (\ie, the network as a whole) or at a local level (\ie, communities or specific nodes), this is not sufficient because it is important to gain insight into the major factors that contribute to the bias. Thus, a framework that supports visual unfairness diagnosis for graph embeddings by explicitly illustrating the sources of bias can be invaluable. A further challenge is that multiple \textit{(un)fairness notions} exist. In particular, these notions can model discrimination either against \textit{groups} \cite{ijcai2019-456} or \textit{individuals} \cite{InFoRM} and it is well-established that trade-offs exist between the two types \cite{kleinberg2016inherent, tradeoff}. Therefore, an integrated design is required that supports heterogeneous fairness notions.

In this work, we propose \biascope, an interactive open-source visualization tool that supports end-to-end \textit{visual unfairness diagnosis} for graph embeddings. To design the tool, we worked with domain experts in order to understand the intricacies of unfairness characterization for graph embeddings. We designed \biascope as a web-based visualization tool with emphasis on interactive features which simplify the complex task of unfairness characterization even on large benchmark networks. Our tool facilitates visual comparison of any two embedding algorithms with respect to fairness on several common network benchmarks. Additionally, it allows the user to locate nodes that contribute to unfairness and importantly, to identify the defining factors of the observed bias. This is facilitated by interactively linking the embedding subspace (\ie, high-dimensional vectors) that displays bias with the corresponding graph topology (\ie, nodes and edges), and vice versa (Figure \ref{fig:teaser}C). Our tool supports both group \cite{ijcai2019-456} and individual \cite{InFoRM} fairness notions in an integrated design and it is task agnostic in the sense that it does not assume a particular downstream task.

The main contributions of our work can be summarized as follows:
\begin{itemize}
    \item Through an iterative interview process we identified the \textit{abstract tasks} that are crucial parts of an \textit{unfairness diagnosis workflow} for graph embeddings. 
    \item Informed by our task analysis, we \textit{designed and implemented} \biascope to support multiple fairness notions, while being interactively configurable and task-agnostic. 
    \item Finally, we collected \textit{expert feedback} that validates our design. The feedback suggests that \biascope can be an effective asset both for researchers to thoroughly evaluate their proposed algorithms and for practitioners to detect bias early in the product development process. 
\end{itemize}

The source code of \biascope is provided at \url{https://github.com/agapiR/BiaScope}, where the reader can download a \textit{demo video} introducing the tool and its functionality. The repository also provides instructions for running a local instance of \biascope.

\section{Related Work}

\biascope lies at the intersection of vector embedding visualization and the analysis of fairness for graphs embeddings. Additionally, we draw inspiration from graph neural network visualization tools.

\paragraph{Vector Embeddings.}

Most vector embedding visualization tools are tailored to contextualized embeddings produced by language models. The tool proposed in \cite{berger2020visually} presents a sentence view with labels for semantic properties e.g., parts-of-speech and an embedding view with cluster analysis via co-occurrence small multiples. On the other hand, embComp \cite{embcomp} introduces visual analysis methods for embedding comparison which combine a global overview with detailed views. The tool mainly focuses on comparing embedding space properties, such as neighborhood overlap or spread and neighbor distances. {The Embedding Comparator system \cite{2022-embedding-comparator} shares this objective, differentiating from embComp in the fact that it simultaneously visualizes global views of embedding structure alongside local views of individual objects and their common and unique neighbors to enable efficient analysis.} Vector embedding comparison is also supported by Emblaze \cite{sivaraman2022emblaze} which consists of an elaborate interactive scatterplot and mainly focuses on neighborhood discovery for dynamic relation suggestions. A more suitable tool for graph embedding visualization is EmbeddingVis \cite{li2018embeddingvis} which allows for comparison of different graph embedding models with respect to which properties of the graph they preserve and illustrates relationships between node metrics and selected embedding vectors.

\paragraph{Graph Neural Networks.}

A common method of learning graph embeddings involves Graph Neural Networks (GNNs), for which there are several visualization tools. The GNNlens tool \cite{gnnlens} helps researchers open the black box of Graph Neural Networks. The tool utilizes Parallel Sets View and Projection View for quick identification of error patterns in the set of wrong predictions. Moreover, a detailed overview for a particular node is offered by Graph View and Feature Matrix View. In a similar spirit, GNNExplainer \cite{gnnexplainer} assists with interpretability by presenting explanations for predictions made by a GNN. The explanations have the form of small visualizable graph motifs and important node features. On the other hand, the CorGIE tool \cite{CorGIE} focuses on the correspondence between the graph topology and the node embeddings. It utilizes a $k$-hop graph layout to show topological neighbors in hops and their clustering structure. It uses small multiples bar charts for node features and a UMAP for the 2D projection of the latent space. Finally, the GNNVis \cite{gnnvis} paper proposes a framework for learned dimensionality reduction using GNNs.

\paragraph{Fairness in Machine Learning.}



Several notions of fairness have been proposed in the literature, mainly categorized into two families: statistical (group) and individual fairness, where most of the focus lies on the first type \cite{chouldechova2020snapshot, 10.1145/3447548.3467266}. This also holds in the context of graph mining, as algorithmic fairness definitions have been adapted for ranking, clustering, and embedding \cite{fair-graph-tutorial}. A graph mining method consists of three major components: the input graph, the mining model and the mining results \cite{InFoRM}. The main idea behind individual fairness is that similar individuals should receive similar algorithmic outcomes. In the context of graph mining, this translates into the fact that “similar” nodes on the graph are also “similar” in terms of the mining results. The similarities between nodes are encoded in a similarity matrix. In particular, every pair of nodes must satisfy the fairness condition. 

Group fairness notions in the context of embeddings have also been studied \cite{pmlr-v97-bose19a, ijcai2019-456}. Bose and Hamilton \cite{pmlr-v97-bose19a} introduce an adversarial framework to enforce fairness constraints on graph embeddings, by training a set of “filters” to prevent adversarial discriminators from classifying the sensitive information from the filtered embeddings. After training, these filters can be composed together in different combinations, allowing for the flexible generation of embeddings that are invariant with respect to any subset of the sensitive attributes. Rahman et al. \cite{ijcai2019-456}, extend the Node2Vec algorithm in a fairness-aware manner. For doing so, they propose a new notion of fairness in the context of friendship recommendation systems, by extending statistical parity.

Algorithmic fairness has also received attention in the visualization community. Several tools that target different machine learning settings (e.g., classification and ranking) have been proposed: FAIRVIS \cite{FAIRVIS} is a visual analytics system that helps audit the fairness of binary classification models. The system allows users to explore both suggested and user-specified subgroups and supports 10 metrics for comparison by default (Accuracy, Recall, Specificity, Precision, Negative Predictive Value, False Negative Rate, False Positive Rate, False Discovery Rate, False Omission Rate, and F1 score). Users can derive new metrics from the base outcome rates. Another tool proposed for the classification setting is DiscriLens \cite{9222272}, which identifies a collection of potentially discriminatory attributes based on causal modeling and classification rules mining and provides visualization to facilitate the exploration and interpretation of them. For high-level model understanding, the What-If Tool \cite{whatif} offers visualizations of selected performance and fairness measures as well as model comparison with respect to those measures. Although the tool provides a comprehensive visual overview of multiple aggregate statistics, it does not offer insights into the sources of the observed bias. Focusing on ranking settings, Fairsight \cite{FairSight}, incorporates different fairness notions (including group and individual fairness) to support identifying and mitigating bias against individuals and groups in problems that involve rank-ordering individuals. Finally, FairRankVis \cite{fairrankvis} enables the exploration of multi-class bias in graph ranking algorithms, allowing for the comparison of fair and unfair versions of these algorithms using both group and individual notions of fairness. FairRankVis is the only tool that supports model comparison as a mechanism for explaining the impacts of algorithmic debiasing. However, its design is tailored to graph ranking since its main visual components target the comparison of two algorithms with respect to their ranking results. In our work, we focus on graph embeddings without restricting our analysis to a particular machine learning application. To the best of our knowledge, \biascope is the first tool that allows unfairness diagnosis for graph embeddings in a task-agnostic manner. 

\section{Preliminaries}

In this section, we discuss the definition and formalism we adopt for \textit{graph embeddings} (Section \ref{sec:graph_emb}). Then, this formalism is used to define the \textit{fairness notions} we consider, for which we also provide intuition and examples (Section \ref{sec:fairness}). Finally, we briefly discuss the data we used for the present design study (Section \ref{sec:data}).

\subsection{Graph Embeddings} \label{sec:graph_emb}

Graph embedding algorithms take a graph $G = \left(V, E\right)$ as input and represent each node in $\mathbb{R}^d$ where $d$ is a dimensionality parameter provided to the algorithm. The node-level vector representations are chosen such that two nodes $v_i$ and $v_j$ that are close to each other in $G$ are embedded near each other in $\mathbb{R}^d$ \cite{hamilton2017representation}. Graph embedding algorithms can also be clustered into several methodological classes: matrix factorization \cite{lapeigenmap, HOPE}, random walk \cite{node2vec}, auto-encoder \cite{SDNE}, and GCNs \cite{graphSAGE}. Recent work has also embedded nodes into hyperbolic space, which better represents hierarchical tree-like structures \cite{HGCN, poincare}. 

In what follows, we assume that the graph embedding is represented by the embedding matrix $Y \in \mathbb{R}^{n \times d}$, where $Y[u] \in \mathbb{R}^d$ is the embedding of node $u$ and $n = |V|$ the number of nodes in the graph.

\subsection{Fairness Notions} \label{sec:fairness}
Recent work has evaluated the fairness of graph embeddings and builds upon broader fairness definitions established in the algorithmic fairness community. Fairness definitions typically fall into two classes: individual and group. Individual fairness ensures that algorithms treat two similar individuals similarly. In contrast, group fairness ensures that two sub-populations, in aggregate, are treated similarly. In this work, we incorporate embedding fairness definitions from both classes. In the following, we provide the formal definitions and illustrative examples.

\subsubsection{Individual Fairness}

\begin{figure}
\centering
    \begin{subfigure}
        \centering
        \begin{tikzpicture}[main/.style = {draw, circle}] 
        \node[main] (1) {A};
        \node[main] (2) [right of=1] {B};
        \node[main] (3) [below of=1]{C};
        \node[main] (4) [right of=3] {D};
        \draw (1) -- (2);
        \draw (1) -- (3);
        \draw (3) -- (4);
        \end{tikzpicture}
    \end{subfigure}
    \begin{subfigure}
      \centering
      \includegraphics[width=.6\linewidth]{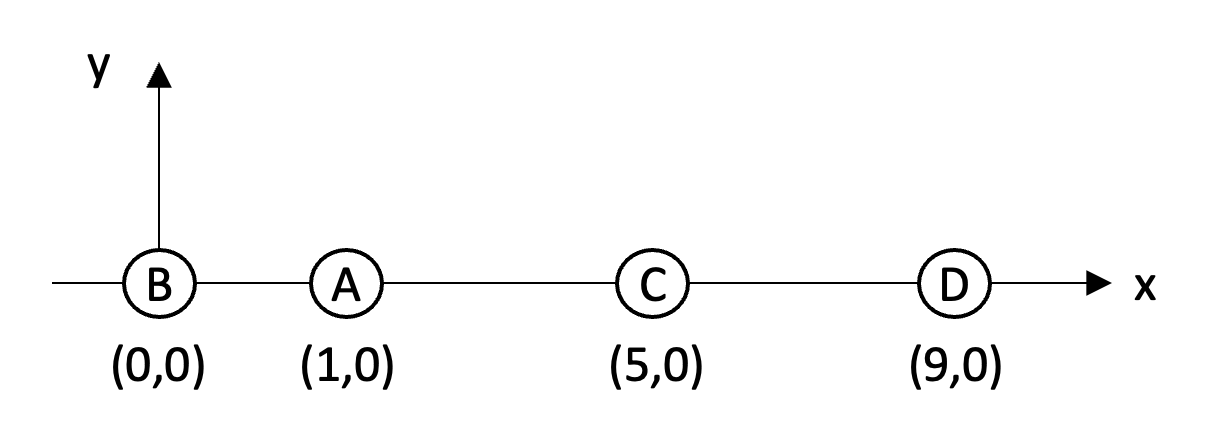}
    \end{subfigure}%
    \caption{Example graph and corresponding embedding for $\text{score}_1$.}
    \label{fig:ex1_graph}
\end{figure}

For individual fairness, we build upon the definition of individual fairness for graph embeddings introduced in \cite{InFoRM} (InFoRM), which assigns a non-negative unfairness score to each node. The score indicates how differently the node is embedded from other nodes in the neighborhood, where the neighborhood is defined by the number of hops from the target node. For example, consider the graph shown in Figure \ref{fig:ex1_graph}. Since nodes B and C are the (1-hop) neighbors of A, the closer their embeddings are to the one corresponding to node A, the lower A's unfairness score. Intuitively,  we expect \textit{similar} nodes according to the local topology of the graph to also be \textit{similar} in terms of their distance in the embedded space. We can formalize this notion by considering a \textit{proximity} or \textit{similarity} matrix $S$ where $S[u,v]$ is the proximity of nodes $u$ and $v$ (i.e., a quantitative measure that indicates how close or similar they are in the graph). The most obvious notion of node proximity is the adjacency matrix of the graph $A$, but other options include its powers ($A,\dots,A^k$), random walk matrices as well as several other neighborhood overlap measures as explained in \cite{grl_book}. Assuming $S=A$ and $Y$ being the embedding matrix, we can define the individual fairness score based on \cite{InFoRM} as follows:
\begin{align*}
    \text{score}_1(u,k=1) &= \sum_{v \neq u} \lVert Y[u] - Y[v]\rVert_2^2 \cdot S[u,v] \\
    &= \sum_{v\in\mathcal{N}(u,1)} \lVert Y[u] - Y[v]\rVert_2^2
\end{align*}
where $\mathcal{N}(u,k)$ is the $k$-hop neighborhood of $u$, given by the nodes reachable from $u$ in at most $k$ steps. In the example from Figure \ref{fig:ex1_graph}, we get $\text{score}_1(\text{A},1)=17$. We generalize this notion for different values of $k$ as follows:
\begin{align*}
    \text{score}_1(u,k) &= \sum_{v\in\mathcal{N}(u,k)} \lVert Y[u] - Y[v]\rVert_2^2
\end{align*}
Returning to our example (Figure \ref{fig:ex1_graph}), considering $k=2$, node D is now taken into account since it is reachable from node A in two hops. Thus, we get $\text{score}_1(\text{A},2)=81$. Observe that for a given vertex $u$ and $k\geq 0$, $\text{score}_1(u,k)\leq \text{score}_1(u,k+1)$ and that isolated nodes get a score of 0. 

The scores for a graph with minimum degree at least one are normalized to be in the range $[0,1]$ as follows:
\begin{align*}
    \text{score}^d_1(u,k) &= \frac{\text{score}_1(u,k)}{\text{degree(u)}}\\
    \text{score}^n_1(u,k) &= \frac{\text{score}^d_1(u,k)}{\max_u \text{score}^d_1(u,k)}
\end{align*}
where $\text{degree(u)}$ is the degree of node $u$ and $\text{score}^n_1(u,k)$ denotes the normalized score.


\subsubsection{Group Fairness}

\begin{figure}
    \centering
    \includegraphics[scale=0.33]{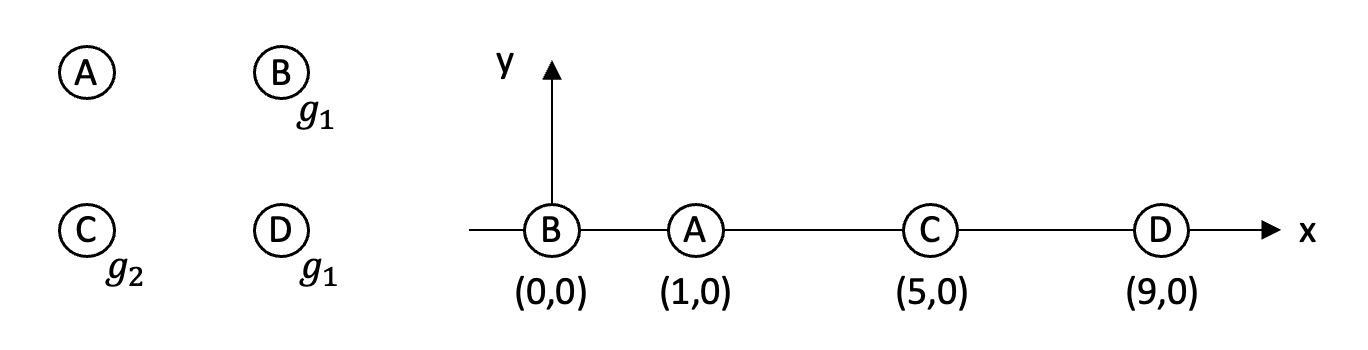}
    \caption{Example graph and corresponding embedding for $\text{score}_2$.}
    \label{fig:ex2_graph_emb}
\end{figure}

Regarding group fairness, our score is based on the definition of group fairness for graph embeddings proposed by \cite{ijcai2019-456} (Fairwalk). The motivation behind it is to promote recommendations (in the context of a Recommender System) where all population groups are equally represented. To start, a set of link recommendations are made for each node. Recommendations are the top $k$ closest embeddings among non-connected nodes according to dot product similarity. We then measure the proportion of recommendations belonging to a specific population (e.g., number of edges recommending men and women). A high score (in magnitude) suggests that one sub-population is recommended disproportionately more. As an example, consider an attribute $S$ with values $z\in\{g_1,g_2\}$ and the disconnected nodes displayed in Figure \ref{fig:ex2_graph_emb}, each annotated with their  attribute value. For $k=2$, the top $2$ most proximal embeddings to the embedding of node A are the ones associated to nodes B and C. In this scenario, both groups (given by the nodes for which $z=g_1$ and $z=g_2$ respectively) are equally represented in the recommendations, which is intuitively what we would identify with being (group) fair. Note that if $k=1$ this situation cannot be achieved.

Formally, given a sensitive attribute $S$ with value $z$ and denoting the set of ``recommended" nodes\footnote{When computing $\rho_k(u)$, a criteria is needed for breaking ties in the ranking, for instance one can use smallest node id goes first.} by $\rho_k(u)$, we can restrict this set based on the attribute value as follows:
\begin{align*}
    \rho_{k,z}(u) &= \{v \;:\; v\in\rho_k(u) \wedge \texttt{attr}_S(v)=z\}.
\end{align*}
Next, defining $Z^S$ to be the set of possible values for attribute $S$ we can measure the fraction of recommended nodes with attribute value $z$ as
\begin{equation}
    \texttt{share}(u,k,z) = \frac{|\rho_{k,z}(u)|}{|\rho_k(u)|} .
\end{equation}
In order to compute how much it deviates from the \textit{equal representation} scenario, we can use the following expression
\begin{equation}
    \texttt{score}_2(u,k,z) = \frac{1}{|Z^S|} - \texttt{share}(u,k,z).
\end{equation}
In the example (Figure \ref{fig:ex2_graph_emb}), for $k=2$ we get the ordered lists $\rho_{2}(A)=\langle B, C \rangle$, $\rho_{2,g_1}(A)=\langle B\rangle$, $\rho_{2,g_2}(A)=\langle C\rangle$ and since $Z^S = \{g_1,g_2\}$, we have $\texttt{score}_2(A,2,g_1) = \texttt{score}_2(A,2,g_2) = \frac{1}{2}-\frac{1}{2}=0$, which corresponds to the \textit{equal representation} scenario. On the contrary, as noted before, if we take $k=1$; $\rho_{2}(A)=\langle B \rangle$, $\rho_{2,g_1}(A)=\langle B\rangle$ and $\texttt{score}_2(A,1,g_1) = \frac{1}{2}-1=-\frac{1}{2}$. Symmetrically, we get $\rho_{2,g_2}(A)=\langle \rangle$ and $\texttt{score}_2(A,1,g_2) = \frac{1}{2}-0=\frac{1}{2}$. Both represent (inevitably) \textit{unbalanced} recommendations among the 2 groups.

\subsection{Data} \label{sec:data}
We use graph embeddings for several commonly used graphs in the graph machine learning community. As previously introduced, graph embedding algorithms represent each node in a graph as a $d$-dimensional vector. So given a graph with $n$ nodes, a graph embedding algorithm outputs an $\mathbb{R}^{n\times d}$ matrix where rows correspond to nodes and columns to graph embedding dimensions. 

\begin{table}[ht]
    \centering
    \begin{minipage}{\linewidth}
    \centering
    \begin{tabular}{| l | l | r | r | r |}\hline
         \textbf{Graph} & \textbf{Type} & $n$ & $m$ & $C$\\ \hline
         Wikipedia \cite{node2vec}& Language & 4.8K & 185K & 1\\
         Facebook \cite{snapnets}& Social & 4.0K & 88K& 1\\
         PPI \cite{node2vec}& Biological & 3.9K & 77K & 35\\
         LastFM \cite{snapnets}& Social & 7.6K & 28K & 1 \\
        \hline
    \end{tabular}
    \end{minipage}
    \caption{We test our design on four real-world graphs spanning multiple domains. All of the networks are thousands of nodes. Above $n$ is the number of nodes, $m$ is the number of edges, and $C$ is the number of connected components.}
    \label{tab:snap_datasets}
\end{table}

The graphs we used are shown in Table \ref{tab:snap_datasets}. All of these graphs are real-world graphs from multiple domains. As group fairness definitions require sensitive-attribute node labels, we retrieved the gender labels that accompany the Facebook network. For anonymity the gender labels are ``0" and ``1". Four of the $4,039$ nodes did not have labels, and we chose to impute a value of ``0" for these 4 missing labels. 

The graph embedding algorithms we use are SVD, HOPE \cite{HOPE}, Laplacian Eigenmaps \cite{lapeigenmap}, Node2Vec \cite{node2vec}, SDNE \cite{SDNE}, and Hyperbolic GCN (HGCN) \cite{HGCN}. These algorithms span the multiple classes of graph embedding algorithms: matrix factorization, random walk, and deep learning. We ran these embedding algorithms on the graphs listed in Table \ref{tab:snap_datasets} and the embeddings are stored in the following repository: \url{https://github.com/dliu18/embedding_repo}. 

\section{Task Analysis for \biascope}

\subsection{Expert Interviews}


In order to understand the key challenges of thoroughly investigating bias in graph embeddings we established a collaboration with Professor Tina Eliassi-Rad and Dr. Brennan Klein. Eliassi-Rad is a Professor of Computer Science and a core member of Northeastern’s Network Science Institute and the Institute for Experiential AI at Northeastern University; she has led research in both of the domains of interest and is familiar with many of the real-world applications. Dr. Klein holds a Ph.D. from Northeastern’s Network Science Institute; he is a network visualization expert and active network science researcher. 

We conducted an initial interview process that led to our task analysis presented in Section \ref{subsec:task_anal}. We divided our interview into three sections: questions related to 1) graph visualization 2) algorithmic fairness and 3) real-world applications of graph embeddings. For each section we prepared a few high-level questions to understand current difficulties in the respective domains. Our interview yielded three main takeaways that inspired our task analysis and design. First, regarding visualizing networks, our domain expert collaborators emphasized the importance of having different visualization encodings for networks of different scales. More specifically, in order to fully characterize bias of graph embeddings it is imperative to allow for both global (whole-network) as well as local (communities, single nodes) views. Second, as most networks are very large to visualize it is important to have summary network statistics to complement the visualization. Additionally, certain graph statistics reflect network properties (\eg, connectivity, degree distribution) that influence how bias in the network is incorporated more by certain graph embedding algorithms compared to others. Third, for algorithmic fairness tools, Professor Eliassi-Rad highlighted the Aequitas system \cite{aequitas} because the interface guides the user through the complicated decision process of choosing fairness definitions and configurations. The takeaway is that a thorough analysis of bias should take into account multiple fairness definitions and configurations and the user should be guided in the process of multifaceted bias analysis. Finally, Dr. Klein connected us to several network visualization libraries \cite{aslak2019netwulf, schwab2021ssvg, wapman2019webweb} that are commonly used by network science researchers for creating customs visualizations, as part of their research process. Testing these recommendations allowed us to understand the domain conventions regarding network visualization and incorporate the most ubiquitous ones in our design. 

\subsection{Task Analysis}\label{subsec:task_anal}

\begin{table*}[ht]
\centering
\label{tab:tasks}
\begin{tabular}{|p{0.2\linewidth}|p{0.3\linewidth}|p{0.07\linewidth}|p{0.07\linewidth}|p{0.08\linewidth}||p{0.08\linewidth}|}
\hline

\multicolumn{2}{|c|}{\textbf{Domain Tasks}} & \multicolumn{4}{|c|}{\textbf{Abstract Tasks}}\\ \hline
Task & Example and Description & High & Mid & Low & Graph-Specific \\\hline

\textbf{Statistical Summary} & Degree distribution, avg. degree, clustering coefficient, triangle count & Derive & Explore & Characterize Distribution & Overview \\\hline

\textbf{Embedding Comparison \textit{w.r.t.} Fairness} & Compare two embeddings w.r.t. a fairness score per node, per topological community or for the whole graph & Present & Explore & Correlate & Overview \\\hline

\textbf{Unfairness Diagnostics} & For a given embedding, find which nodes contribute to unfairness and why & Discover & Locate & Find Anomalies & Attribute-based: On the Nodes\\\hline

\end{tabular}
\caption{\biascope is designed to support the domain tasks described in the table, which facilitate our complete unfairness diagnosis workflow for graph embeddings. For each domain task on the left, we provide a detailed explanation and then we list the high, mid, and low level abstract tasks \cite{munzner2014visualization, amar2005low} along with the graph-specific abstract task \cite{lee2006task}.}
\end{table*}

From the interaction with our domain expert collaborators we identified the key components of a typical unfairness diagnosis approach for graph embeddings. In particular, we first extracted and ranked the high level domain tasks mentioned by our interviewees within the unfairness diagnosis procedure. We then constructed concrete examples for each activity. This analysis led us to select the most significant domain tasks our tool is designed to support. Subsequently, we mapped the domain tasks to abstract tasks that can be addressed by our visualization design. 

The \textit{task analysis} we conducted is outlined in Table \ref{tab:tasks}.  We refer to the identified domain tasks by title and augment each domain task with informative descriptions, listed under the ``Example and Description'' column of Table \ref{tab:tasks}. For the abstract task extraction, we used Munzer’s Actions Taxonomy \cite{munzner2014visualization} for the High and Mid levels, and the Analytic Task Taxonomy proposed by Amar et al. \cite{amar2005low} for the Low level. We augmented the abstract task analysis with graph-specific tasks from the task taxonomy proposed by Lee et al. \cite{lee2006task}. 

We designed \biascope to support three main tasks: 
\begin{itemize}
    \item[\textbf{T1}] Obtain a comprehensive \textit{Statistical Summary} of a network chosen by the user. The summary conveys structural properties possibly related to bias, i.e., degree distribution, edge density, etc.
    \item[\textbf{T2}] \textit{Compare} two embeddings of the network w.r.t. a fairness score. The two embeddings are computed using two different algorithms. The fairness score is chosen and configured by the user.
    \item[\textbf{T3}] \textit{Diagnose} the observed unfairness of an embedding. This task allows the user to transition from a global to a local perspective, by focusing on the subgraph and embedding subspace that together determine the unfairness score of a user-selected node.
\end{itemize}

The outlined tasks facilitate a typical unfairness diagnosis workflow for graph embeddings. The workflow consists of an \textit{overview} step and a \textit{drill-down} step:
\begin{enumerate}
    \item Overview: Comparison between two embedding algorithms (\textbf{T2}), augmented by relevant graph statistics (\textbf{T1}). In this step, the user examines whether the embedding of interest displays signs of unfairness, compared to other embeddings of the same network. 
    \item Drill-down/Diagnosis: Identification of the unfairness source(s) for the selected embedding (\textbf{T3}). In this step, the user discovers the determining factors as well as the nature of the observed bias, by focusing on certain nodes or communities that are embedded most unfairly. 
\end{enumerate}

\section{Design of \biascope} \label{sec:design}



This section outlines and justifies the selected visual encoding and interactions implemented in \biascope. The selection was performed by analyzing a series of independently drawn sketches that were created to satisfy the tasks identified from our Task Analysis (Section \ref{subsec:task_anal}). In what follows, a high level description of the different encodings is given. Specific details related to the particular network being used in the examples is deferred to Section \ref{sec:usage_scenario}.

We satisfy the statistical summary task (\textbf{T1}) with the ``Statistical Summary of the Network''\footnote{We also refer to this component as the \textit{Statistical Summary View}.} component of the \textit{Overview}, shown in Figures \ref{fig:teaser}A and \ref{fig:statistical-summary}. The table encoding containing the summary provides key graph metrics to help the user understand the type of network being studied. For the Degree Distribution the bar chart idiom was selected, since it makes use of the most effective magnitude channel to encode ordered attributes, according to the effectiveness ranking for visual channels presented in \cite{munzner2014visualization}. The ranking was compiled by Munzner based on previous work performed by the Visualization community such as \cite{cleveland_mcgill_84a, cleveland_93a, mackinlay_86, Ware12, heer_bostock_10} .

\begin{figure*}[t]
    \centering
    \includegraphics[scale=0.4]{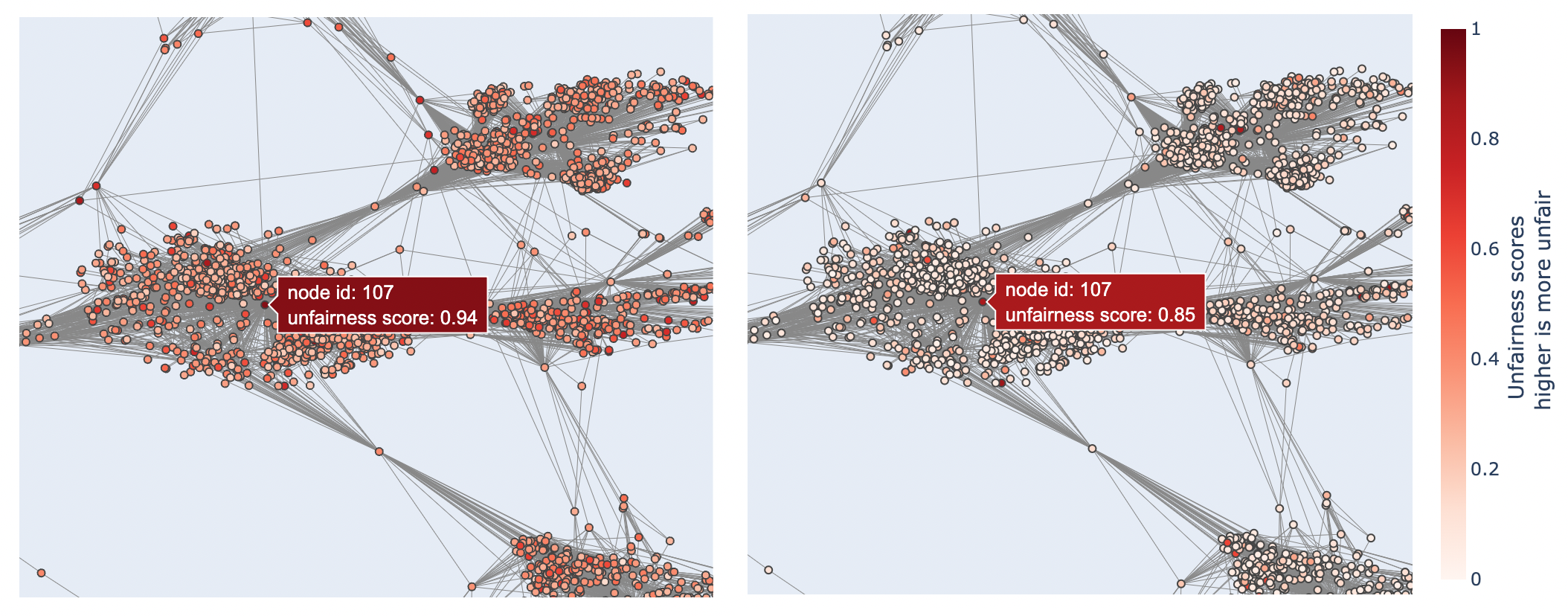}
    \caption{Node attributes (ID and associated fairness score) are displayed on mouse hover. Zooming functionality is also supported to inspect and compare the scores of specific parts of the network.}
    \label{fig:mouse_hover}
\end{figure*}

Next, we achieve the comparison task (\textbf{T2}) by providing side-by-side views in the \textit{Overview}, whose main visual components\footnote{We also refer to this component as the \textit{Comparison View}.} are shown in Figures \ref{fig:statistical-summary} and \ref{fig:inform-diagnose}. In particular, the same network is displayed on both sides of Figure \ref{fig:inform-diagnose} using the spring layout \cite{fruchterman1991graph}, while the node colors depend on the fairness of the respective embedding algorithm. A sequential color scale is used, since the attribute being encoded is quantitative. We associate darker colors with higher unfairness scores due to their negative nature \cite{bartram2017affective}. 
Regarding interactivity, the node id and its fairness score are displayed upon mouse hover (Figure \ref{fig:mouse_hover}) for the user to be able to better asses the difference between two nodes of the network. Zooming is  supported to allow for better analysis capabilities of specific network communities (Figure \ref{fig:mouse_hover}). 

Due to the large scale nature of our data, following \cite{munzner2014visualization}, we have implemented visual feedback in the form of a loading sign while the different visualizations are being loaded into the webpage. Further details regarding data management and implemented preprocessing steps are provided in Section \ref{sec:preprocessing}. 

\begin{figure*}
    \centering
    \includegraphics[scale=0.39]{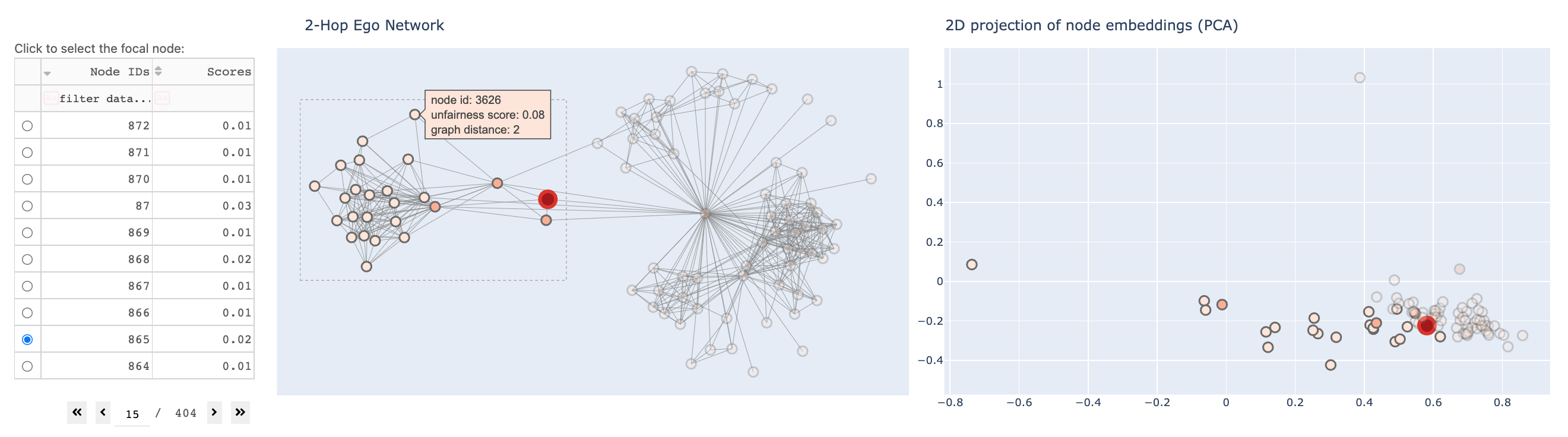}
    \caption{Main visual components of the \textit{Diagnose View} for individual fairness. From left to right: Interactive table listing node ids and their unfairness scores, local neighborhood of selected node in the network and corresponding projected embeddings.}
    \label{fig:diagnose_view_1}
\end{figure*}

Finally, we address the unfairness diagnostics task (\textbf{T3}) with the \textit{Diagnose View}. This view lists the nodes together with their unfairness scores, allowing the user to select a focal node and uncover the cause(s) of bias. Specifically, the main components of the view are: (1) an interactive table containing the node ids along with their unfairness scores, (2) the projected embeddings affecting the score of the focal node, and (3) the corresponding subgraph topology. The user can sort the list of nodes based on their ids or scores by clicking on the arrows next to the column names of the Table (left of Figure \ref{fig:diagnose_view_1}). A search box is included for each attribute of the table to speedup the lookup. 

As previously discussed, we consider two fairness notions (individual and group) with different characteristics, which are reflected by our design. For the case of individual fairness, the score of a node is determined by its local neighborhood. Thus, the local neighborhood topology along with the corresponding projected embeddings are displayed in the view. On the other hand, group fairness determines the score of a node by the most proximal embeddings. Therefore, these projected embeddings are shown along with the corresponding subgraph. In both cases, upon selection of the focal node, the \textit{Diagnose View} shows side-by-side the subgraph and projected embeddings relevant to the chosen fairness notion. Figure \ref{fig:diagnose_view_1} displays the \textit{Diagnose View} for the case of individual fairness. The projected embeddings\footnote{The projection of the embeddings is performed using PCA \cite{PCA_FRSLIIIOL}. In what follows, for simplicity, we do not account for the projection when referring to the projected embeddings.} are encoded using a scatterplot (Figure \ref{fig:diagnose_view_1}, rightmost chart), noting that by \cite{munzner2014visualization} it maximizes effectiveness. 

\begin{figure}
    \centering
    \includegraphics[scale=0.31]{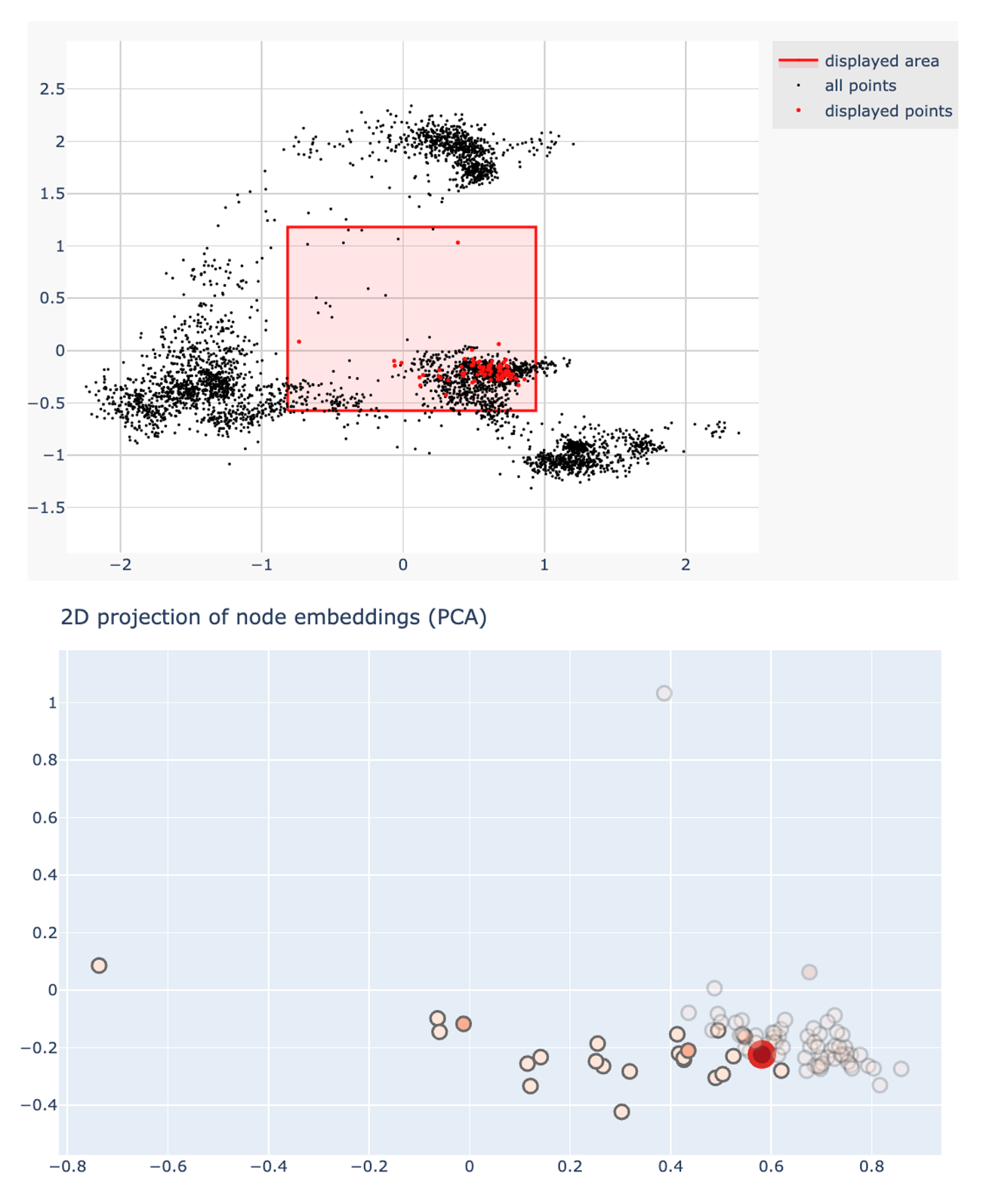}
    \caption{The context legend displayed in the \textit{Diagnose View} (top plot) conveys the scale of the distances observed in the scatterplot (bottom plot), where the embeddings that affect the score are plotted.}
    \label{fig:diagnose_view_2}
\end{figure}

The objective of the \textit{Diagnose View} is to provide insight into the fairness score computation for the selected focal node and uncover causes of unfairness in the network. For this purpose, a red color pop-out effect is used to outline the focal node, together with an increase in its size. This improves the efficiency of its lookup \cite{munzner2014visualization}, which is an essential component in understanding the score computation. Brushing and linking is supported between the network and the embedding space to analyze how different nodes contribute to the fairness score of the focal node. Specifically, when brushing over a set of elements, the corresponding elements in the other view are highlighted. This is shown in Figure \ref{fig:diagnose_view_1}, where a subset of nodes in the local topology of node with id $865$ is being selected with the brush. When hovering over a node or its embedding, the corresponding node id and fairness score are displayed in the tooltip, together with the number of hops to the focal node for the individual fairness score (Figure \ref{fig:diagnose_view_1}) or the node label for group fairness. The distance from the focal node (individual fairness) and the gender (group fairness) are encoded using the color channel with an ordinal and categorical color scale, respectively. 
Furthermore, as a response to expert feedback, we improved our design with a context legend that conveys the scale of the distances observed in the scatterplot (Figure \ref{fig:diagnose_view_2}). More specifically, we made use of the \textit{focus+context} approach \cite{cockburn_etal_08}, and augmented the \textit{Diagnose View} (i.e., \textit{focus}) with a global scatterplot (i.e., \textit{context}) displaying all the embeddings (top section of Figure \ref{fig:diagnose_view_2}), where the highlighted area corresponds to the points of the scatterplot which are the ones that affect the focal node (Figure \ref{fig:inform-diagnose} and \ref{fig:fairwalk-diagnose}).

\section{Evaluation}

\subsection{Usage Scenario}\label{sec:usage_scenario}
\begin{figure}
    \centering
    \includegraphics[width=\linewidth]{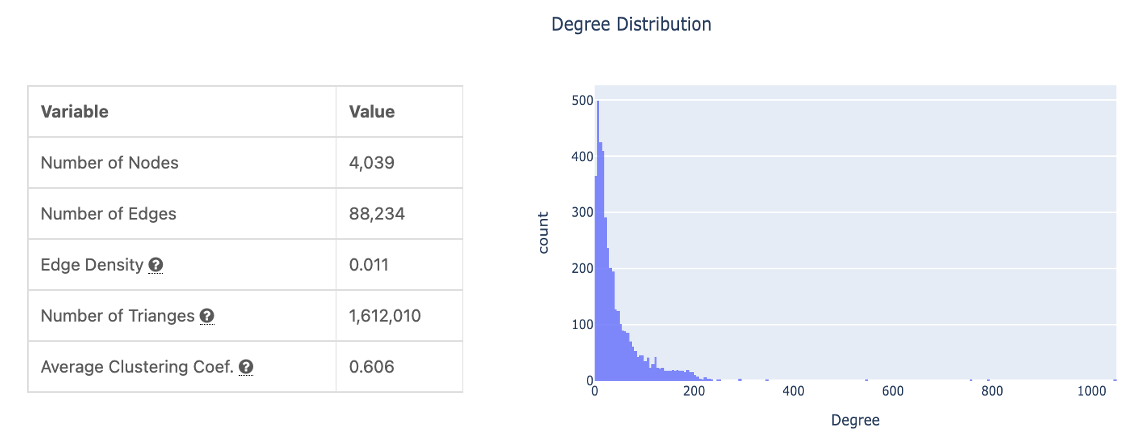}
    \caption{The statistical summary for the Facebook network helps the user to characterize the data before analyzing its fairness scores. The table on the left states that the network is sparse and highly clustered while the degree distribution shows that almost all users in the network have fewer than 200 friends.}
    \label{fig:statistical-summary}
\end{figure}

To illustrate the design in action, we will walk through a case study using the Facebook network, which represents the social network friendships of over four thousand users. We begin with an overview of the network via the statistical summary shown in Figure \ref{fig:statistical-summary}. The table on the left informs the user that the Facebook graph is sparse and highly clustered. Further, the degree distribution shows that almost all users in the network have fewer than 200 friends.

Proceeding further into the \textit{Overview}, the user can browse a side-by-side comparison of the fairness scores for two sets of embeddings. Figure \ref{fig:teaser}B shows the scores for Node2Vec on the left and then HGCN on the right for the Facebook network. This visual provides two takeaways. First, the Node2Vec embeddings are overall more fair than the HGCN embeddings, as indicated by the lighter coloring. Second, the unfairness for HGCN is concentrated in select communities. Without the visualization, a purely statistical analysis would resort to aggregate values which do not account for heterogeneity. 

\begin{figure}
    \centering
    \includegraphics[width=\linewidth]{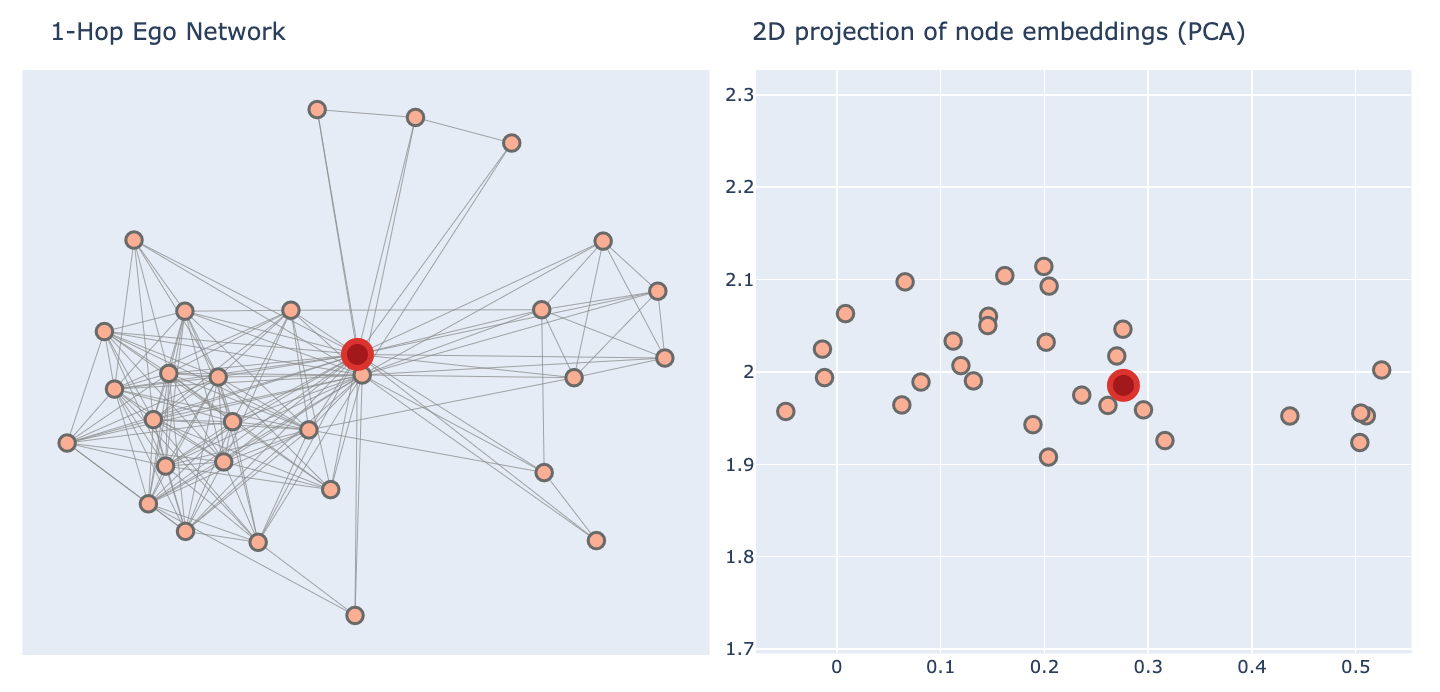}
    \caption{Moving into the drill-down tasks, the above figure provides the diagnostics results for the Facebook network's Node2Vec embeddings where fairness is defined with the InFoRM notion of individual fairness. The ego network on the left shows the target node is part of two communities, as linked on the right hand side, these communities are themselves embedded far apart. Thus, the diagnostics results show that under InFoRM bridge nodes between communities may be embedded unfairly. }
    \label{fig:inform-diagnose}
\end{figure}

Moving into the drill down portion of the tasks, we can now use the \textit{Diagnose View} to better understand why certain nodes are scored as unfair. For instance, Figure \ref{fig:inform-diagnose} shows the diagnostic results for the Facebook Node2Vec embeddings using the individual fairness notion. The local neighborhood (ego network) shows that the focal node, in red, is part of two communities of friends. Further, the projected embeddings show that the two communities of friends are embedded far from each other. Hence, the focal node is scored as unfair because it is far from its neighbor communities in embedding space. This mapping between neighbors and embeddings can be verified with brushing and linking (Figure \ref{fig:diagnose_view_1}). 

\begin{figure}
    \centering
    \includegraphics[width=\linewidth]{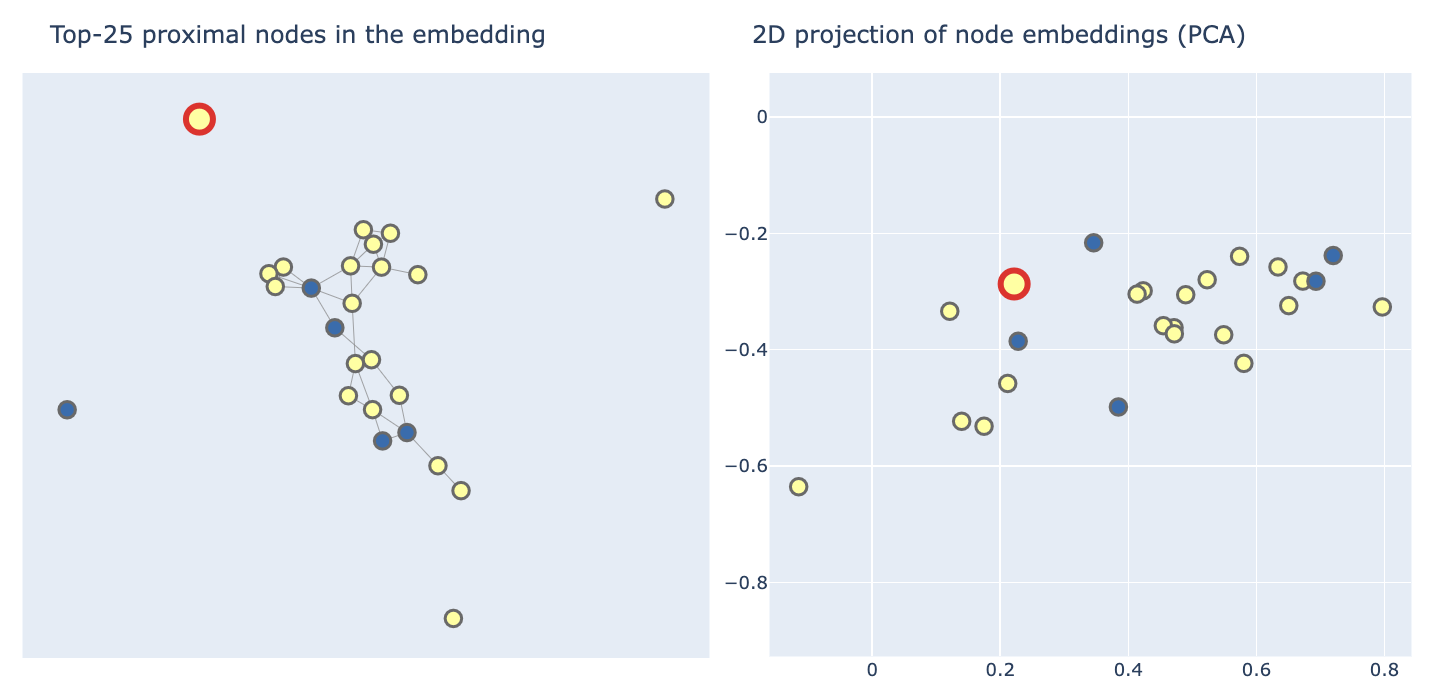}
    \caption{The diagnostics results for group fairness show that the given node from the Facebook network is unfair because many of its recommended nodes are also of the same gender value, encoded in yellow. The focal node and the induced subgraph given by the top-$k$ recommended nodes are plotted, together with the corresponding embeddings. The design layout is similar to the individual fairness diagnostics but the node colors now represent the sensitive attribute, which is the most important attribute for group fairness.}
    \label{fig:fairwalk-diagnose}
\end{figure}

Finally, the user can also drill-down and diagnose the group fairness scores as well. Figure \ref{fig:fairwalk-diagnose} shows the Facebook Node2Vec embeddings evaluated based on the group fairness notion. Now, the focal node together with the induced subgraph (top-$k$ recommended nodes) is plotted in the rightmost chart. The subgraph shows that the target node's recommendations are mostly of gender 0, encoded in yellow, which is a form of homophily. Further, these nodes are embedded close to the target node in the embedding space. This is a form of group unfairness because if we train a model to recommend friends based on these embeddings, the model would perpetuate homophily for this node and recommend nodes of the same gender. 

\subsection{Expert Feedback}\label{sec:expert_review}

In order to evaluate our tool, we conducted feedback interviews with our two collaborators and two additional experts, who are active researchers interested in graph embeddings, also affiliated with Northeastern’s Network Science Institute. We conducted four individual interviews, during which each participant was able to interact with the tool through the webpage on their own system. We simultaneously observed their interaction with the tool, which was captured on video for further analysis. For all interviews we followed the same steps: First, we asked for the participant's consent to video recording. Second, we briefly discussed the tasks our tool is designed to support. Third, we provided an \textit{interactive walk-through} of the tool by asking the participant to locate certain views and perform concrete tasks. The goal of this step was to help the participants familiarize themselves with the UI. Next, we encouraged the participant to follow a \textit{usage scenario} similar to the one presented in Section \ref{sec:usage_scenario}. During this step we provided loose guidance by setting open-ended goals and we encouraged the participant to freely accomplish them using the tool. Lastly, we collected verbal feedback using the following questions:
\begin{itemize}
    \item[\textbf{Q1}] In your view, how well are the domain tasks supported?
    \item[\textbf{Q2}] How would you perform these tasks without the tool? Do you see value in our tool?
    \item[\textbf{Q3}] How would you use our tool in your research/development process?
    \item[\textbf{Q4}] What limitations do you identify?
\end{itemize}
Following the interview sessions, we collected additional quantitative feedback from the experts using a System Usability Scale (SUS) \cite{sus_1}. In what follows, we summarize the key takeaways from the feedback. 

\paragraph{Effectiveness and Usability.} 



During the interactive walk-through, we observed that all the participants were able to quickly perform the tasks with little to no assistance. Regarding explicit verbal feedback, all experts agreed that \biascope effectively supports the described domain tasks (\textbf{Q1}). Additionally, all participants described, while three explicitly said, that they would follow the same process in order to thoroughly analyse unfairness of a graph embedding (\textbf{Q2}). Most participants pointed out that without \biascope this analysis would be tedious and prone to errors. One participant mentioned that \biascope would save them time and a different one said: ``something that goes a long way in terms of what you have done is I do not want to implement any of these things [by myself]. ... at the very basic level you have just reallocated a lot of work... [Since there are no other tools like this one on the market], this is immediately useful in a very time saving way.''.

Regarding the potential role of \biascope within our participants' research process (\textbf{Q3}), the feedback validated the tool's effectiveness while additionally providing new perspectives. Two of the experts agreed that the tool would facilitate the evaluation of their graph embedding algorithms with respect to fairness. More specifically, they mentioned the following: ``For people who aren't really fairness focused, this would [enable] a nice quick sanity check.'' and ``I would use it as a visual tool to interrogate the data, [which is] a very important thing to do, as opposed to just feeding it into your machine learning system and then hoping for the best.'' The other two experts pointed out they could further utilize our tool to facilitate their work in explainable machine learning or for pedagogical purposes:  ``... I would definitely use this tool for understanding explainability and interpretability of embeddings ... my major use would be like explaining some of the down stream task results or predictions'', ``It is a very good teaching tool, in general I think web based [tools] that let you interact with networks, change the game when it comes to learning about networks.''.

\paragraph{Limitations.}

The experts agreed on the importance of an ``import" feature, which would allow the user to upload and diagnose their own embedding algorithm. One of the experts additionally suggested an ``export" functionality. The implementation of an “import” feature is included as future work in Section \ref{sec:limitations_futurework}. Moreover, some experts pointed out that the Diagnose View should include an indication of the scale regarding the projected embeddings, in order to clearly convey the magnitude of distances observed in the local embedding subspace. We addressed this limitation with the addition of the context legend, as discussed in Section \ref{sec:design}. One of the experts expressed the interesting idea to augment our network options with certain simple synthetic networks representing typical graph topologies, e.g., star, which could enlighten the user with respect to the nature of different fairness notions. Finally, adding on-demand information about the selected fairness notion definition was a frequent suggestion among participants.

\paragraph{System Usability Scale.} 
We requested our experts to fill a System Usability Scale (SUS) \cite{sus_1, brooke2013sus}, a standard simple ten-item scale giving a global view of subjective assessments of the usability of a system. Based on the responses, we derive the corresponding SUS scores, which have a range of 0 to 100. Regarding scores assessment, \cite{bangor_etal} proposed an adjective value scale for SUS scores. According to this scale, scores in the range 72.5-84.9 are associated with a good usability and above 85 represent an excellent one. Based on the collected SUS scores, we obtained a mean value of $83.125$. This implies, per the aforementioned scale, that \biascope's usability was perceived as good, close to excellent by the interviewed experts.




\section{Discussion}

In this section, we discuss several points that were observed during the development process and posterior usage of \biascope. These consist of challenges regarding the size of the data, both in terms of visibility and the efficiency of the system (Section \ref{sec:preprocessing}), as well as limitations and future lines of work, included in Section \ref{sec:limitations_futurework}.

\subsection{Data Management}\label{sec:preprocessing}
Due to the size of the networks that were considered, several performance challenges needed to be addressed in the development process. In this section, we provide an overview of how these were approached.

As previously outlined, in order to improve network visibility, we filter out visually non-salient edges. Specifically, we only display edges in the top 10 percent of length. Filtering allows us to save on browser memory, as the number of edges far exceeds the number of nodes. At the same time, by inspecting the networks we found that the bottom 90 percent of shortest edges were often not visible, so filtering offers large performance increases while preserving important visual information, e.g., intra-community connectivity.

In terms of preprocessing, both the projection of the embeddings, as well as group and individual fairness scores were calculated offline in order to obtain an additional performance speedup. For the later, this involves generating for every network the corresponding score for each node id and possible configuration. Configuration options consist of the number of hops from the focal node for the individual fairness notion, while for the group fairness, the value of $k$ as well as the chosen sensitive attribute (in our case gender) and value.

Finally, the pre-computed network information is accessed only once when the webpage is initially loaded and stored in memory, in order to avoid redundant server access.

\subsection{Limitations and Future Work}\label{sec:limitations_futurework}


In the present design study we handled large benchmark networks, as outlined in Section \ref{sec:preprocessing}, therefore we had to overcome responsiveness issues. As previously discussed, we used a simple heuristic to minimize the number of edges in our visualization by filtering out non-salient edges. In future work, we plan to explore more sophisticated heuristics as well as different methods for graph abstractions e.g., sub-sampling methods that preserve node communities. Additionally, we plan to extend our scope to a larger suite of benchmark networks which are even larger (in terms of node count) or have higher density. A parallel direction of future work includes developing advanced interactivity and linking in the overview, e.g. simultaneous zooming, coupled hover events, etc. We expect this extension to generate new performance challenges since the interaction should be particularly responsive in order to increase effectiveness. 

Furthermore, following the expert feedback, we plan to make certain additions in order for the tool to better support the needs of its target users. These include the extension of our benchmark suite with other networks popular within the graph mining community (e.g., Books \footnote{\url{http://www-personal.umich.edu/~mejn/netdata/}}, NS \cite{netscience-data}) and a list of different options for the embedding projections beside PCA (e.g., UMAP \cite{becht2019dimensionality}). Finally, as mentioned in Section \ref{sec:expert_review}, we plan to develop a user friendly onboarding process for users who want to evaluate their own embedding algorithms using our tool. The onboarding will include preprocessing of the user data, which will utilize server-side computation, in order to preserve sufficient responsiveness at use-time.

\section{Conclusion}

Motivated by the recent efforts made by the algorithmic fairness community, we have build \biascope, an interactive web-based visualization tool that supports end-to-end \textit{visual unfairness diagnosis} for graph embeddings. Our design is the result of an iterative collaborative process with experts in the fields of fair machine learning and graph mining. To inform our design, we conducted a thorough task analysis that consists of key tasks that support an unfairness diagnosis workflow for graph embeddings.

Applying our visual unfairness diagnosis workflow to a benchmark network, we show the effectiveness our tool has in visually revealing the unfairness incorporated into different widely used graph embedding algorithms. Our work is part of a greater effort to allow both researchers and practitioners to build or use machine learning models while simultaneously being mindful of bias and algorithmic fairness, without the need of expertise in the field. Our tool, and others in the same spirit, could be part of a standard development workflow of a machine learning researcher or engineer, as it is designed to provide greater transparency into how current graph embedding algorithms and fairness notions interconnect in practice.

\section {Acknowledgement}
The authors would like to thank Zohair Shafi and Ayan Chatterjee for participating in the expert feedback sessions and providing insightful comments about the system.

\newpage 

\bibliographystyle{abbrv-doi}
\bibliography{bibliography}

\end{document}